\documentclass[fleqn,usenatbib]{mnras}

\usepackage{newtxtext,newtxmath}
\usepackage[T1]{fontenc}
\usepackage{ae,aecompl}
\usepackage{graphicx}	% Including figure files
\usepackage{amsmath}	% Advanced maths commands
\usepackage{amssymb}	% Extra maths symbols
\usepackage{multicol}        % Multi-column entries in tables
\usepackage{bm}		% Bold maths symbols, including upright Greek
\usepackage{pdflscape}	% Landscape pages
\usepackage{soul}
\usepackage{enumitem}
\usepackage{pdflscape}	% Landscape pages
\usepackage[dvipsnames]{xcolor}
\newcommand{\Myr}{\mathrm{Myr}}

\newcommand{\kpc}{\mathrm{kpc}}
\newcommand{\pc}{\mathrm{pc}}
\newcommand{\kmpers}{\mathrm{km} \, \mathrm{s}^{-1}}
\newcommand{\cmcube}{\mathrm{cm}^{-3}}
\newcommand{\cmsquare}{\mathrm{cm}^{-2}}
\newcommand{\protonmass}{m_\mathrm{p}}
\newcommand{\mppercmcube}{\protonmass \, \cmcube}

\newcommand{\Msol}{\textup{M}_\mathrm{\sun}}

\newcommand{\tenMsol}[1]{10^{#1} \,\Msol}
\newcommand{\xMsol}[2]{\ensuremath{{#1}\times 10^{#2} \,\Msol}}
\newcommand{\xScientific}[2]{\ensuremath{{#1} \times 10^{#2}}}
\newcommand{\xScientificErrorBar}[4]{\ensuremath{{#1}^{+#2}_{-#3} \times 10^{#4}}}
\newcommand{\Msolyr}{\textup{M}_\mathrm{\sun} \, \text{yr}^{-1}}

\newcommand{\mdm}{m_{\mathrm{DM}}}

\newcommand{\Mvir}{M_{200}}

\newcommand{\Mstar}{M_{\star}}
\newcommand{\rvir}{r_{200}}

\newcommand{\rhalflight}{r_{1/2}}

\newcommand{\nh}{n_{\mathrm{H}}}
\newcommand{\nhi}{n_{\mathrm{\hi}}}
\newcommand{\xhi}{x_{\mathrm{\hi}}}
\newcommand{\Mhi}{M_{\mathrm{\hi}}}
\newcommand{\rhi}{r_{\mathrm{\hi}}}
\newcommand{\NhibelowXX}[2]{\Nhi \leq \xScientific{#1}{#2}\, \cmsquare}
\newcommand{\NhiaboveXX}[2]{\Nhi \geq \xScientific{#1}{#2}\, \cmsquare}
\newcommand{\rhiaboveXX}[2]{r(\NhiaboveXX{#1}{#2})}
\newcommand{\Nhi}{N_{\mathrm{\hi}}}
\newcommand{\hinospace}{H\textsc{i}}
\newcommand{\hi}{H\textsc{i} }
\newcommand{\hmol}{\mathrm{H}_\mathrm{2}}

\newcommand{\MstarMhi}{\Mstar - \Mhi}

\graphicspath{{./}{figs/}}

\title[Scatter in \hi observables of faint dwarfs]
{EDGE: What shapes the relationship between \hi and stellar observables in faint dwarf galaxies?}

\author[M. P. Rey et al.]{Martin P. Rey,$^{1, 2}$\thanks{E-mail: \href{martin.rey@physics.ox.ac.uk}{martin.rey@physics.ox.ac.uk}} Andrew Pontzen,$^{3}$  Oscar Agertz,$^{2}$ Matthew D. A. Orkney,$^{4}$ Justin I. Read,$^{4}$
\newauthor Am\'elie Saintonge,$^{3}$ Stacy Y. Kim$^4$ and Payel Das$^{4}$
\vspace{0.8mm}
\\
$^{1}$ Sub-department of Astrophysics, University of Oxford, DWB, Keble Road, Oxford OX1 3RH, UK \\ 
$^{2}$ Lund Observatory, Department of Astronomy and Theoretical Physics, Lund University, Box 43, SE-221 00 Lund, Sweden \\
$^{3}$ Department of Physics and Astronomy, University College London, London WC1E 6BT, UK\\
$^{4}$ Department of Physics, University of Surrey, Guildford GU2 7XH, UK
}

\date{Accepted 2022 February 19. Received 2022 February 3; in original form 2021 December 6}

\pubyear{2021}

\begin{document}
\label{firstpage}
\pagerange{\pageref{firstpage}--\pageref{lastpage}}
\maketitle

\begin{abstract}
We show how the interplay between feedback and mass-growth histories introduces scatter in the relationship between stellar and neutral gas properties of field faint dwarf galaxies ($\Mstar \lessapprox \tenMsol{6}$). Across a suite of cosmological, high-resolution zoomed simulations, we find that dwarf galaxies of stellar masses $10^5 \leq \Mstar \leq \tenMsol{6}$ are bimodal in their cold gas content, being either \hinospace-rich or \hinospace-deficient. This bimodality is generated through the coupling between (i) the modulation of \hi contents by the background of ultraviolet radiation (UVB) at late times and (ii) the significant scatter in the stellar-to-halo-mass relationship induced by reionization. Furthermore, our \hinospace-rich dwarfs exhibit disturbed and time-variable neutral gas distributions primarily due to stellar feedback. Over the last four billion years, we observe order-of-magnitude changes around the median $\Mhi$, factor-of-a-few variations in \hi spatial extents, and spatial offsets between \hi and stellar components regularly exceeding the galaxies' optical sizes. Time variability introduces further scatter in the $\MstarMhi$ relation and affects a galaxy's detectability in \hi at any given time. These effects will need to be accounted for when interpreting observations of the population of faint, \hinospace-bearing dwarfs by the combination of optical and radio wide, deep surveys. 
\end{abstract}

%%%%%%%%%%%%%%%%%%%%%%%%%%%%%%%%%%%%%%%%%%%%%%%%%%

%%%%%%%%%%%%%%%%% BODY OF PAPER %%%%%%%%%%%%%%%%%%

\section{Introduction} \label{sec:intro}

Mapping the cold gas content of galaxies and its relationship with their stellar observables provides a unique view into the physics of the interstellar medium (ISM) and the regulation of star formation in galaxies (e.g. \citealt{Young1995, Bigiel2008, Catinella2010, Saintonge2011, Saintonge2017,  Emsellem2021} and references therein). This has recently become possible for the faintest galaxies, as the combination of large radio surveys mapping \hi emission (e.g. \citealt{Haynes2011, Peek2011}) and deep follow-up photometric imaging has uncovered a population of \hinospace-bearing, often star-forming, low-mass dwarf galaxies ($\Mstar \leq 10^7\, \Msol$; \citealt{Irwin2007, Cole2014}; \citealt{McQuinn2015, McQuinn2020, McQuinn2021}; \citealt{Sand2015, Adams2018, Brunker2019, Janesh2019, Hargis2020, Bennet2022}). 

The shallow potential wells of such faint systems make them particularly sensitive to feedback processes that regulate star formation (see \citealt{Somerville2015, Naab2017} for reviews), originating both from within their ISM and from the population of galaxies throughout the Universe. In particular, explosions of massive stars are efficient at driving outflows in dwarf galaxies (e.g. \citealt{Dekel1986, Christensen2016}), but their efficiency in removing cold gas and modulating \hi emission in such systems remains an open question (e.g. \citealt{Agertz2020EDGE, Smith2020PhotoRT}). In addition, once the intergalactic medium has been heated by the ultraviolet (UV) radiation of the population of galaxies and quasars around a redshift of $z\approx6$, its Jeans' mass is raised accordingly, preventing gas accretion onto the lowest-mass dark matter haloes (e.g. \citealt{Efstathiou1992, Noh2014}, see \citealt{McQuinn2016} for a review). Without gas inflows, the faintest dwarfs are unable to maintain a significant, cold ISM, eventually shutting down their star formation activity even up to the present time (e.g. \citealt{Ricotti2005, Hoeft2006, Onorbe2015, Benitez-Llambay2017, Agertz2020EDGE}). This suppression thus couples uniquely with each dwarf's past history, depending on their respective dynamical masses at the time of reionization (e.g. \citealt{Okamoto2008, Benitez-Llambay2015, Fitts2017,Rey2019UFDScatter}). Furthermore, at later times ($z\leq2$), the dwindling of star formation and quasar activity coupled with cosmological expansion steadily reduce the strength of the UVB ($z\leq1$; \citealt{McQuinn2016}). This late decay as each object continues to grow in dynamical mass is key in regulating any re-accretion of gas and potential re-ignition of star formation in dwarf galaxies (\citealt{Ricotti2009,Benitez-Llambay2017, Ledinauskas2018, Rey2020, Benitez-Llambay2021}). The coupling between these time-dependent feedback mechanisms and each dwarf's history is thus expected to introduce significant diversity in their observed stellar and gaseous properties.

The prospects of discovering numerous \hinospace-rich dwarfs in the Local Volume in the near future provide us with a timely opportunity to quantify the scatter in \hi contents of gas-rich dwarfs and constrain these feedback mechanisms. Next-generation imaging surveys will reach depths capable of resolving faint systems in the field (e.g. the Vera Rubin Observatory, \citealt{Ivezic2019}; the Euclid Space Telescope, \citealt{Scaramella2021}). This will allow systematic cross-matching with current catalogues of \hi clouds (e.g. \citealt{Adams2013}) as well as those from ongoing and future radio surveys (e.g. MeerKAT, \citealt{Maddox2021}; WALLABY, \citealt{Koribalski2020}; Apertif; \citealt{vanCappellen2022}; the Five-Hundred-Meter Aperture Spherical radio Telescope, \citealt{Zhang2021FASTExtragalacticHI}; the Square Kilometer Array; \citealt{Braun2019}). Performing such cross-matching is highly desirable, as pinpointing which dark matter haloes retain measurable \hi contents and host active star formation has strong constraining power on alternative dark matter models (e.g. \citealt{Papastergis2011, Pontzen2012, DiCintio2014, Nadler2021MWConstraints}, see \citealt{Pontzen2014, Bullock2017} for reviews) and cosmic reionization and its associated UV suppression (e.g. \citealt{Ricotti2005, Tollerud2018, Benitez-Llambay2020}).

In this work, we use a suite of high-resolution, zoomed cosmological simulations from the `Engineering Dwarf Galaxies at the Edge of galaxy formation' (EDGE) project (first presented in \citealt{Agertz2020EDGE}) to study how feedback mechanisms shape the diversity of \hi contents in the faintest galaxies. Zoomed, hydrodynamical simulations of galaxy formation allow us to capture the cosmological mass assembly of a handful of faint dwarf galaxies and provide an approximate treatment for large-scale radiative effects while affording sufficient resolution to resolve the ISM of such small objects (e.g. \citealt{Maccio2017, Revaz2018, Munshi2019, Rey2019UFDScatter, Wheeler2019, Agertz2020EDGE, Applebaum2021, Gutcke2021CosmologicalSim, Orkney2021}). Each of our simulations has sufficient resolution to resolve individual supernova explosions within a dwarf galaxy's ISM, strongly reducing numerical uncertainties in the injection of energy and momentum from supernovae (e.g. \citealt{Kim2015, Martizzi2015}). EDGE simulations also make use of the genetic modification technique (\citealt{Roth2016, Rey2018, Stopyra2021}), allowing us to create alternative cosmological histories for a chosen dwarf galaxy while maintaining all its other cosmological aspects, such as environment. Such an approach is ideal to gain physical insights into the interplay between feedback and mass growth, enabling a causal account of how it shapes the observables of faint dwarfs (\citealt{Rey2019UFDScatter, Rey2020, Orkney2021}). 

We describe our simulation suite and how each galaxy is evolved to $z=0$ in Section~\ref{sec:methods}. We present their neutral gas properties in Section~\ref{sec:results} focusing on the mechanisms generating scatter in the relationship between stellar and \hi contents in the population of faint dwarfs. We will provide further insights into the resolved observables and cold gas kinematics of individual dwarfs in a follow-up paper. We summarize and discuss the implications of our findings in Section~\ref{sec:conclusion}.

\section{A suite of simulated faint dwarf galaxies} \label{sec:methods}

\begin{table*}
  \centering
  \caption{Summary of the EDGE suite of low-mass, faint dwarf galaxies. Individual simulated objects (first column) were first described in \citet{Rey2019UFDScatter, Rey2020} and \citet{Orkney2021}, and we reproduce here their $z=0$ dynamical, stellar masses and projected half-light radii (second, third and fourth columns respectively). We quantify in this work their neutral gas properties, computing their \hi masses within the central kpc, their \hi cylindrical extents above a depth of $10^{19} \cmsquare$ and the spatial offsets between their stellar and \hi centres (fifth, sixth and seventh columns respectively). Due to time variability, \hi properties are reported through their median and fifty per cent confidence interval over the last four billion years of each dwarf's evolution.} 

     \begin{tabular}{l c c c c c c c }
     \hline
     Simulation name & $\Mvir$ ($\Msol$)& $\Mstar(<1 \, \kpc)$ ($\Msol$) & $\rhalflight$ ($\pc$) & $\Mhi(<1 \, \kpc)$ ($\Msol$) & $\rhi(\Nhi\geq 10^{19} \, \cmsquare)$ ($\pc$) &$\Delta_{\star-\hi}$ ($\pc$) \\
     \hline

     \textcolor{MidnightBlue}{Halo 600}& $\xScientific{3.3}{9}$ & $\xScientific{3.5}{5}$ & 167 & \xScientificErrorBar{2.9}{3.5}{2.8}{5} & $482^{+166}_{-482}$ & $140^{+105}_{-32}$ \\[3pt]

     \textcolor{RawSienna}{GM: delayed mergers} & $\xScientific{3.2}{9}$ & $\xScientific{3.1}{5}$ & 201 & \xScientificErrorBar{1.9}{2.1}{1.2}{5} & $307^{+112}_{-75}$ & $72^{+34}_{-32}$ \\

     \hline

     \textcolor{MidnightBlue}{Halo 605}& $\xScientific{3.2}{9}$ & $\xScientific{1.6}{6}$ & 206 & \xScientificErrorBar{3.1}{1.3}{0.57}{5} & $456^{+65}_{-55}$  & $87^{+70}_{-45}$ \\

     \hline

     \textcolor{RawSienna}{Halo 624}& $\xScientific{2.5}{9}$ & $\xScientific{5.6}{5}$ & 264 & \xScientificErrorBar{2.1}{0.66}{0.77}{5} & $295^{+44}_{-30}$ & $14^{+4}_{-3}$ \\[3pt]

     \textcolor{MidnightBlue}{GM: higher final mass} & $\xScientific{3.7}{9}$ & $\xScientific{1.2}{6}$ & 233 & \xScientificErrorBar{2.2}{1.38}{0.79}{5} & $359^{+33}_{-98}$ & $84^{+13}_{-36}$ \\

     \hline

     Halo 1459 & $\xScientific{1.4}{9}$ & $\xScientific{2.3}{5}$ & 196 & $<100$ & -- & -- \\[3pt]

     GM: Earlier & $\xScientific{1.4}{9}$ & $\xScientific{4.9}{5}$ & 131 & $<100$ & -- & -- \\[3pt]

     GM: Later & $\xScientific{1.4}{9}$ & $\xScientific{1.1}{5}$ & 273 & $<100$ & -- & -- \\[3pt]

     GM: Latest & $\xScientific{1.4}{9}$ & $\xScientific{2.3}{4}$ & 822 &  $<100$ & -- & -- \\

     \hline

     Halo 1445& $\xScientific{1.3}{9}$ & $\xScientific{7.2}{4}$ & 164 & $<100$ & -- & -- \\

     \hline
     \end{tabular}
   \label{table:runs}
\end{table*}

The simulations used in this study were first presented in \citet{Rey2019UFDScatter, Rey2020} and \citet{Orkney2021}. We briefly describe in Section~\ref{sec:sec:numerics} how their initial conditions are evolved to $z=0$ through zoomed, cosmological simulations (see \citealt{Agertz2020EDGE} for a comprehensive description) and summarize our suite of objects in Section~\ref{sec:sec:suite}.

\subsection{Numerical setup} \label{sec:sec:numerics}

We construct cosmological initial conditions using the \textsc{genetic} software (\citealt{Stopyra2021}) with cosmological parameters $\Omega_{M} = 0.309$, $\Omega_{\Lambda} = 0.691$, $\Omega_{b} = 0.045$, $H_{0} = 67.77 \,\kmpers \text{Mpc}^{-1}$ (\citealt{PlanckCollaboration2014}) and generate zoomed initial conditions reaching dark matter particles masses of $\mdm = 960 \, \Msol$ (see \citealt{Agertz2020EDGE} for further details). We evolve these initial conditions using linear theory to $z=99$ (\citealt{Zeldovich1970}), and then perform cosmological, zoomed galaxy formation simulations to follow the evolution of dark matter, stars, and gas to $z=0$ with the adaptive mesh refinement code \textsc{ramses} (\citealt{Teyssier2002}). Our refinement strategy ensures a spatial resolution of $3 \, \pc$ throughout the galaxy's ISM (\citealt{Agertz2020EDGE}).

We complement this hydrodynamical setup with an extensive galaxy formation model described in detail by \citet{Agertz2020EDGE} as `Fiducial'. We track the cooling of a primordial plasma using hydrogen and helium equilibrium thermochemistry (\citealt{Courty2004, Rosdahl2013}) and model heating from reionization through a spatially uniform, time-dependent UVB based on an updated version of \citet{Haardt1996} as implemented in the public \textsc{ramses} version (see \citealt{Rey2020} for further details). We account for the boost in neutral fraction due to self-shielding at gas densities $\nh \geq 0.01 \cmcube$  (\citealt{Aubert2010, Rosdahl2012}), and star formation proceeds stochastically following a Schmidt law (\citealt{Schmidt1959, Rasera2006, Agertz2013}) in gas with density $\rho \geq 300 \, \mppercmcube$ and temperatures $T \leq 100$ K. 

This modelling and resolution allow us to track the cold and hot phases within the ISM up to star-forming densities, removing uncertainties associated with tracking $\xhi$ through subgrid models of the ISM (e.g. \citealt{Diemer2018}). To compute \hi observables, we thus directly extract the gas \hi fraction, $\xhi$ at every spatial location by evaluating the code's internal cooling function for the local gas temperature and density assuming collisional equilibrium and a primordial hydrogen mass fraction of 0.76. Our model does not explicitly track the formation and destruction of molecular hydrogen in cold gas, potentially leading to unphysically high \hi densities at which gas should transition from \hi to $\hmol$ ($\nhi \gtrapprox 10 \, \cmcube$ and $\Nhi \gtrapprox 10^{21} \, \cmsquare$). We verified that including or excluding such high-density \hi gas has negligible impact on our presented results, and leave to future work a more explicit treatment of molecular chemistry (e.g. \citealt{Christensen2012, Nickerson2018, Sillero2021}).

Combined with this multiphase ISM, our model tracks the injection of energy, momentum, mass, and metallicity according to the progenitor-dependent evolutionary time-scales of stars within a stellar particle (\citealt{Agertz2011, Agertz2013, Agertz2020EDGE}). We account for stellar winds from massive and asymptotic giant branch (AGB) stars, together with the explosions of Type II and Type Ia supernovae. Our numerical resolution is a key aspect of the simulations, allowing us to inject each discrete, individual supernova explosion as thermal energy into the ISM, and self-consistently follow the build-up of momentum by solving the hydrodynamics equations (\citealt{Agertz2020EDGE}). Such a scheme greatly reduces uncertainties in modelling the injection of feedback from supernovae (\citealt{Kim2015,Martizzi2015}).

We identify dark matter haloes using the \textsc{hop} halo finder (\citealt{Eisenstein1998}) and determine the build-up of mass and merger trees of galaxies using the \textsc{pynbody} (\citealt{Pontzen2013}) and \textsc{tangos} (\citealt{Pontzen2018}) libraries by matching haloes between 100 simulation snapshots equally spaced in scale factor (every $\approx 150 \, \Myr$). For each galaxy, we determine the respective centre of stellar, gas and dark matter components using the shrinking sphere algorithm (\citealt{Power2003}) and all projected quantities are determined against a random line of sight unless otherwise stated.

\subsection{Suite summary} \label{sec:sec:suite}

We now present the galaxies used in this work. The EDGE suite builds from five independent dark matter haloes within a narrow range of final dynamical masses ($\xMsol{1}{9} \lessapprox \Mvir \lessapprox \xMsol{3}{9}$; Table~\ref{table:runs}), selected for their high likelihood to host faint dwarf galaxies (e.g. \citealt{Jethwa2018, Read2019SFRMatching, Nadler2020}). All hosts are embedded within a cosmic void, and are isolated field systems, with no neighbours more massive within $5 \, \rvir$ (see \citealt{Orkney2021} for a visual). 

To probe the response of dwarf galaxies to their past histories, three hosts are complemented by genetically modified re-simulations. Genetic modifications make targeted adjustments to the cosmological initial conditions of a given dwarf, in order to achieve a desired modification of their non-linear mass assembly (\citealt{Roth2016, Rey2018, Stopyra2021}). Each modified initial condition makes minimal changes to other untargeted aspects, for example maintaining the same large-scale filamentary structure around the galaxy (e.g. \citealt{Pontzen2017, Rey2019VarianceDMOs}). 

Specifically, we use three genetically modified mass growths targeting the early mass assembly of a low-mass host (Halo 1459 in Table~\ref{table:runs}; see \citealt{Rey2019UFDScatter} for details). These histories sweep through the possible range of dynamical masses at the time of reionization, while all converging to the same final mass at $z=0$. We further include the genetically modified version of a middle-mass halo (Halo 624) increasing its dynamical mass at all times, and the alternate history of a high-mass host (Halo 600) delaying its late ($z\leq1$) mass growth (see \citealt{Rey2020} for details). Combining independent histories with their genetically modified counterparts, we obtain a suite of 10 low-mass, isolated dwarf galaxies whose properties are summarized in Table~\ref{table:runs}.

\section{Results} \label{sec:results}

We now analyse our suite of 10 low-mass, isolated dwarf galaxies with independent histories, which span an extended range of stellar masses ($\xMsol{4}{4} \lessapprox \Mstar \lessapprox \xMsol{2}{6}$; Table~\ref{table:runs}) and exhibit a diversity of gas contents and star formation activity at $z=0$ (\citealt{Rey2020}). To help the analysis of their \hi observables, we order them into three broad classes:
\begin{enumerate}[label=(\roman*)]
  \item \textbf{Gas-poor relics}. They formed the entirety of their stars early ($z\geq4$) and have not retained a sizeable gas reservoir, either hot or cold, by $z=0$ (Halo 1445, Halo 1459 and its modified histories; black in Table~\ref{table:runs}). 
  \item \textbf{Gas-rich relics}. They formed the entirety of their stellar content early ($z\geq4$), but have grown sufficiently in dynamical mass at late times to accumulate gas into their centre (e.g. \citealt{Ricotti2009, Benitez-Llambay2020, Rey2020}). However, the gradual cooling of gas is hampered by internal heating from old stellar populations within the galaxy, and has not yet led to the re-ignition of star formation at $z=0$ (\citealt{Rey2020}; Halo 624 and Halo 600-GM: later mergers\footnote{Although this galaxy has re-ignited star formation at $z=0.03$ ($500 \, \Myr$ ago) to form two generations of stars, its evolution in the last billion years is dominated by slow gas accretion, hence why we place it in this category.}; brown in Table~\ref{table:runs}).
  \item \textbf{Gas-rich, star-forming dwarfs}. They have restarted the formation of stars after a temporary quenching following the reionization of the Universe, and have been actively star forming in the last billion years at average rates $\approx 10^{-5} \, \Msolyr$ (Halo 600, Halo 605 and Halo 624-GM: higher final mass; blue in Table~\ref{table:runs}). As we will see in Section~\ref{sec:sec:hi}, these galaxies can temporarily quench and lose their ISM following feedback cycles, but are able to re-accrete gas and form new stellar populations at least every 400 Myr. 
\end{enumerate} 

\subsection{Neutral gas properties of faint dwarf galaxies} \label{sec:sec:hi}

\begin{figure*}
  \centering
    \includegraphics[width=\textwidth]{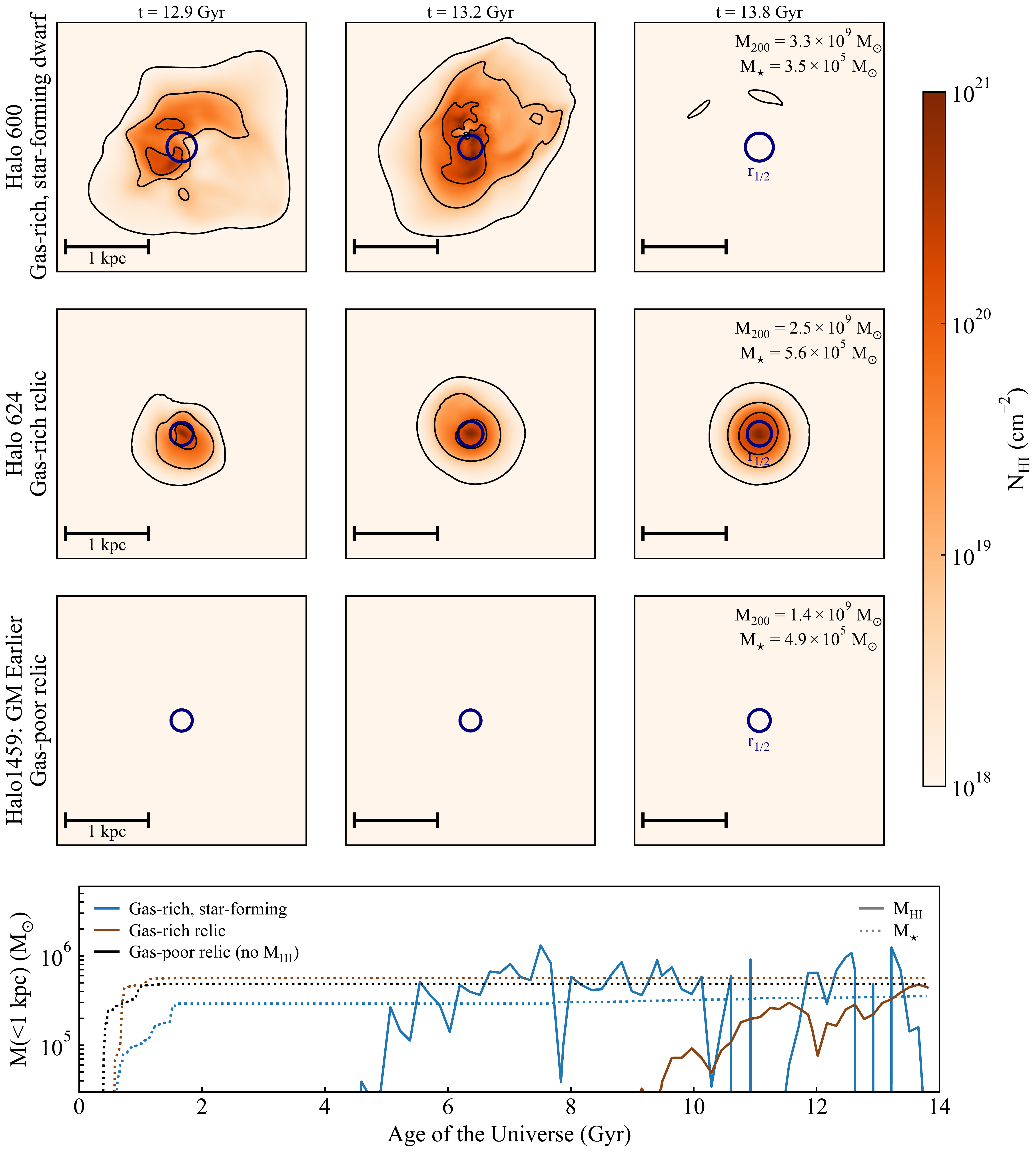}

    \caption{Illustrating the diversity of \hi properties across our suite of simulated low-mass, faint dwarf galaxies, showing neutral hydrogen column density maps at (top to third row; spatial resolution of 6 pc) and \hi content (bottom row) over time for three example objects. Faint dwarfs that have stopped forming stars after reionization can exhibit (i) undetectable levels of \hi (third row and black in the bottom panel) or (ii) sizeable, stable \hi reservoirs if they have grown sufficiently in dynamical mass at late times to re-accrete gas (second row and brown). Galaxies that have further grown enough to re-ignite the formation of young stars (top row and blue) show \hi properties varying over short timescales due to star formation and its associated feedback (left- to right-hand panels), driving variability in their overall content (Figure~\ref{fig:himass}) and spatial structure (Figure~\ref{fig:hisize}). This diversity is generated at similar final galaxy stellar masses at $z=0$ (top right-hand corners, and growth histories in the bottom panel).
    }
    \label{fig:himaps}

\end{figure*}

We start by illustrating the diversity in \hi column densities and masses across our suite of simulated faint dwarf galaxies in Figure~\ref{fig:himaps}. We show the evolution of the central \hi mass (bottom panel) for example objects in our three classes of dwarf galaxies, and plot \hi column density maps (top to third row) at three different snapshot times in the last billion years of evolution (left to right). All images are centred on the stellar component with the projected half-light radius shown as a blue circle. Frames are $3 \, \kpc$ wide, oriented face-on with respect to the gas angular momentum and with a spatial resolution of $6\, \pc$ (i.e. our simulations' resolution, see Appendix~\ref{app:resolution} for alternate spatial resolutions). Contours of constant $10^{18}$, $10^{19}$ and $10^{20} \cmsquare$ \hi column densities are shown in black. 

We start with an example gas-poor, relic dwarf galaxy (black in the bottom panel and third row of panels), which shows undetectable levels of \hi at late times. Such galaxies are in fact almost entirely devoid of gas, both hot or cold in their centres (\citealt{Rey2020}), due to the build-up of the UVB around $z\sim6$ following reionization (see \citealt{McQuinn2016} for a review). This suppresses gas accretion onto low-mass dark matter haloes (e.g. \citealt{Efstathiou1992, Hoeft2006, Noh2014}), preventing gas reservoirs from being replenished after the last star formation events and leaving these galaxies with vanishing \hi contents at $z=0$.

By contrast, dwarf galaxies that assemble a sufficiently deep potential well at late times can overcome this barrier, and re-accrete gas into their centres (e.g. \citealt{Ricotti2009, Benitez-Llambay2020, Rey2020}). This gradual re-accretion might, however, not reach star-forming densities by $z=0$, giving rise to gas-rich, quiescent relics (brown in the bottom panel and second row) and gas-rich, star-forming dwarfs (blue in the bottom panel and top row). Both formation scenarios exhibit extended neutral gas reservoirs at detectable levels for deep, radio observatories ($\NhiaboveXX{5}{19}$; e.g. \citealt{Bernstein-Cooper2014, Adams2018}). 

Quiescent dwarfs have stable \hi content and morphology over time (second row, left to right), while star-forming objects show strong variability (top row). Star-forming dwarfs undergo ejective feedback events following the formation of new stars, driving disturbed \hi morphologies that strongly differ from one timestamp to the next (e.g. top left and centre) and spatial offsets between \hi and stellar components (e.g. top left-hand panel). Furthermore, when observed after an intense outflow, these galaxies may be found without any detectable neutral gas (e.g. top right-hand panel). 

The interplay between each galaxy's dynamical mass growth, photoionization feedback from the cosmic UVB, and the regulative cycle of star formation in a shallow potential well thus leads to a complex diversity in the \hi observables of faint dwarfs at $z=0$. This diversity occurs at nearly constant galaxy stellar mass (top right-hand corners of panels), highlighting strong scatter at the faint end of the $\Mstar$-$\Mhi$ relation, which we quantify next.

\subsection{The $\Mstar-\Mhi$ relation of low-mass dwarf galaxies} \label{sec:sec:mstarmhi}

\begin{figure}
  \centering
    \includegraphics[width=\columnwidth]{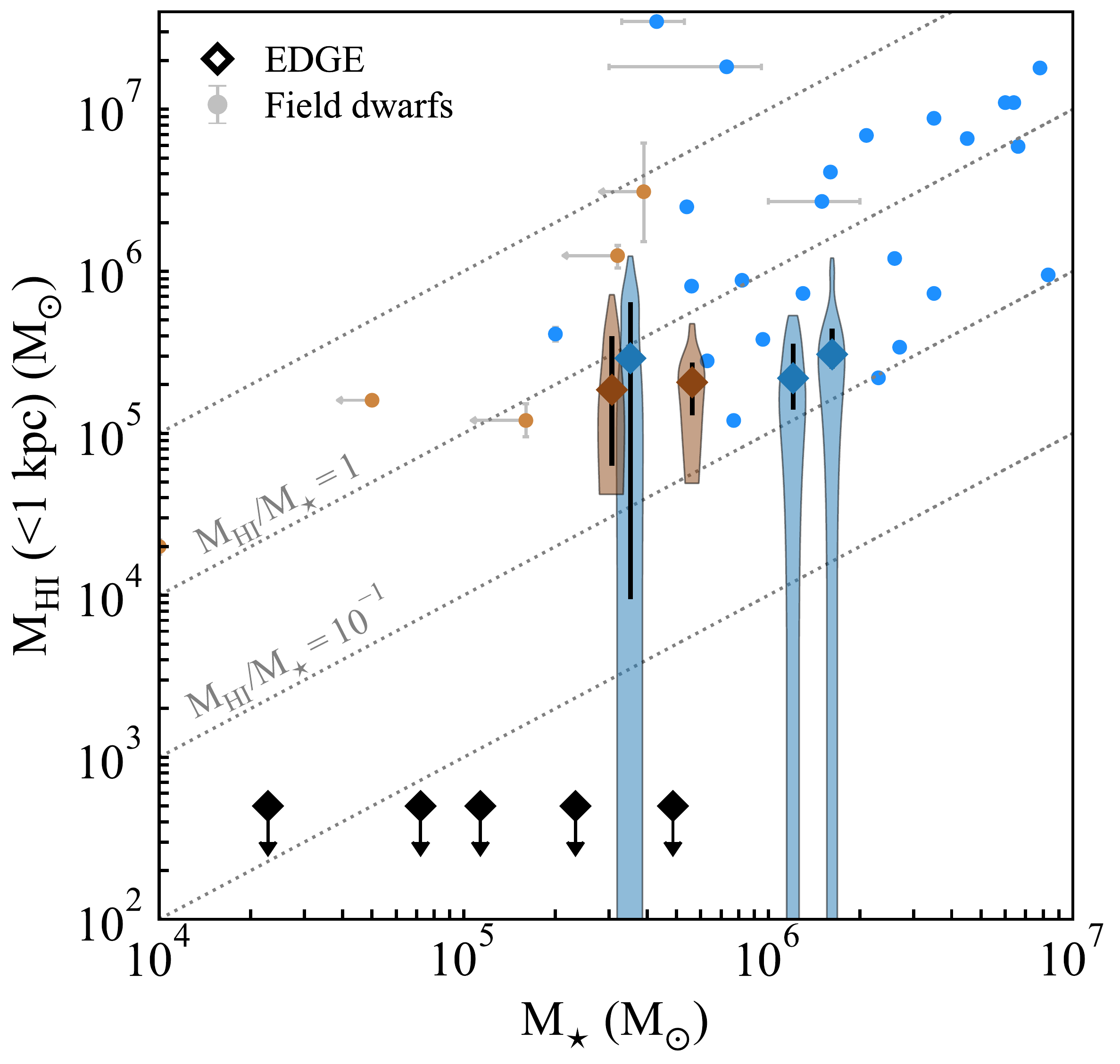}

    \caption{Central neutral hydrogen mass as a function of galaxy stellar mass in our simulated suite. For gas-rich galaxies, variability over time generates an extended distribution in \hi content over the last four billion years (coloured contours show sample densities and black lines show interquartile range), with medians (diamonds) overlapping with the observed population (grey symbols). Feedback episodes can temporarily remove or ionize neutral gas, generating vertical scatter at fixed $\Mstar$ in all \hinospace-rich galaxies, with the most violent events leading to vanishing \hi masses (vertical stripes) in actively star-forming objects (blue), which are absent in quiescent objects (brown). Simulated galaxies that have not retained gas following the reionization of the Universe have undetectable neutral gas contents (black upper limits), but nonetheless overlap in galaxy stellar mass with \hinospace-bearing dwarfs, leading to a bimodal $\MstarMhi$ relation at $z=0$.}

    \label{fig:himass}

\end{figure}

We now quantify the relationship between stellar and neutral gas content for each galaxy in our suite, showing in Figure~\ref{fig:himass} their neutral gas and stellar masses enclosed within $1\, \kpc$. This radius ensures a complete account of \hi and stellar material, enclosing the typical extent of the \hi distribution (Section~\ref{sec:sec:histructure}) and several half-light radii (Table~\ref{table:runs}). As \hi masses are variable over time (Figure~\ref{fig:himaps}), we construct their timeseries over the last four billion years ($z\approx0.5$; $\approx 40$ simulation snapshots)\footnote{We choose this time interval as a compromise between having sufficient snapshot number to obtain statistical properties, while remaining sufficiently small compared to physical evolution within galaxies (e.g. due to the re-accretion of gas; \citealt{Rey2020}).} along the major progenitor and show the sample density (contours), median (diamonds), and the interquartile range (IQR; black lines) of each timeseries. Galaxies without \hi are shown as upper limits (black diamonds).

For comparison, we plot an observed population of Local Volume, field \hinospace-rich faint dwarfs (\citealt{McConnachie2012, Cole2014}; \citealt{McQuinn2015, McQuinn2020, McQuinn2021}; \citealt{Sand2015, Adams2018, Brunker2019, Janesh2019, Hargis2020, Bennet2022}; blue and brown for quiescent and star-forming objects respectively, reporting error bars when available). Median $\Mhi$ for our \hinospace-bearing dwarfs are within the observed population scatter at their stellar masses. [Structural stellar properties also span across their respective observed scatter; \citealt{Rey2019UFDScatter, Rey2020}]. We omit gas-poor faint dwarfs observed as satellites of the Milky Way from this comparison, as it remains unclear how environmental processing affects their \hi reservoirs compared to our simulated field systems. The lack of observed counterpart to our \hinospace-poor faint dwarfs (black diamonds) likely reflects the current observational difficulties to detect such isolated objects in the field without pre-selection on e.g. \hi content, but hints of them might already exist (e.g. Eridanus 2; \citealt{Simon2021}) and their systematic detection should be possible with the next generation of wide-sky deep photometric surveys (\citealt{Simon2019}).

We recover across our whole suite the dichotomy in \hi content illustrated in Figure~\ref{fig:himaps}, with half of our simulated galaxies exhibiting detectable \hi contents, while the other half have no \hi reservoirs. In low-mass haloes, retaining gas is at first order determined by assembling a sufficiently deep potential well at late times, after cosmic reionization (e.g. \citealt{Benitez-Llambay2015, Fitts2017, Benitez-Llambay2020, Rey2020}). If able to retain its gas, a galaxy's average content is then set by its dynamical mass (\citealt{Benitez-Llambay2020}). The similar median \hi contents across our \hinospace-bearing dwarfs is thus naturally explained by their relatively narrow spread in final dynamical masses at $z=0$ (Table~\ref{table:runs}). 

We thus re-affirm that photoionization feedback from the cosmic UVB yields a natural truncation to the faint end of the \hi mass function (e.g. \citealt{Benitez-Llambay2017, Tollerud2018}). However, we demonstrate that this truncation is not associated with a single cut in stellar mass, with dwarfs with $\xScientific{2}{5} \leq \Mstar \leq \xMsol{5}{5}$ exhibiting both $\Mhi \geq \xMsol{1}{5}$ and vanishing $\Mhi$ in our suite. This scatter arises from the interplay between each galaxy's growth of dynamical mass and reionization-induced feedback, in setting both (i) the final stellar mass and (ii) the final \hi content. Unlike the gas content that links to late mass assemblies, the final stellar mass is strongly shaped by how much mass is assembled early, pre-reionization (\citealt{Rey2019UFDScatter}). Sufficient early growth can grow $\Mstar$, without being met by enough late-time growth to create a sizeable \hi reservoir. Conversely, sufficient late growth can lead to measurable $\Mhi$ without being met by early growth, diminishing $\Mstar$. Across a spectrum of final halo masses, the diversity of early and late assemblies hence leads to a bimodality at the faint end of the $\Mstar-\Mhi$ relation, with \hinospace-bearing and \hinospace-poor isolated dwarf galaxies in the same range of stellar masses. 

Our limited sample size does not allow us to robustly quantify the width of this bimodality, which would require modelling a larger population of mass accretion histories and halo masses. However, observational hints already point towards a more extended range of stellar masses, with \hinospace-bearing dwarfs proposed below $\Mstar \leq 10^5 \, \Msol$ (\citealt{Janesh2019}, brown points shown as upper limits in stellar masses). This makes this feature a crucial modelling point for future studies aiming to constrain reionization and cosmological dwarf formation with a population of \hi galaxies (e.g. \citealt{Tollerud2018}).

Beyond the bimodality observed across our suite, we also note that each \hinospace-rich dwarf exhibits significant spread in neutral gas mass across their past four billion years, with 50 per cent confidence interval (black lines) around the median (diamonds) ranging from 0.2 to 1.8 dex. These variations are generated at nearly fixed stellar masses, which hardly evolve over the same time-scale. [Stellar mass loss in old stellar populations is small, and the formation of young stars at rates $\approx 10^{-5} \, \Msolyr$ in active galaxies contributes at most a few per cent of their total stellar mass; \citealt{Rey2020}]. Time variability thus provides an additional physical mechanism to generate scatter in the $\Mstar-\Mhi$ relation of low-mass dwarfs. 

The physical nature of time variability is distinct for star-forming (blue) and quiescent dwarfs (brown). Quiescent dwarfs see their content vary due to the combination of (i) slow accretion of gas increasing their total gas reservoir, and (ii) feedback from old, pre-reionization stars such as Type Ia supernovae and AGB stars stirring up their internal reservoirs (\citealt{Rey2020}). By contrast, dwarfs that are actively forming stars undergo violent events, following the explosion of Type II supernovae in newborn massive stars. This leads to a marked difference in their respective time distributions, with active galaxies showing systematic long tails towards low or vanishing \hi contents (blue vertical stripes). Furthermore, despite its extended extent, our uncovered regular time variability in isolated field systems is unlikely to explain the full breadth of the population scatter. Environmental interactions with nearby structures or mergers provide a way to trigger sudden, single replenishments of the ISM and subsequent increases in $\Mhi$ (e.g. \citealt{Wright2019}), which would add to our mechanisms and might help explain extremely \hinospace-rich objects (e.g. Coma P; \citealt{Brunker2019}; blue circles with $\Mstar$/$\Mhi > 10$ marked by a grey line).

The partial or total removal of neutral gas by short-lived, feedback-driven events can thus affect our ability to detect \hi emission from such galaxies at a given time. We quantify this duty cycle in our three star-forming galaxies, finding that each galaxy spends 3, 4 and 20 per cent of the last four billion years with $\Mhi \leq 10^2 \, \Msol$ (i.e. no neutral gas reservoir), and 3, 22 and 34 per cent with $\Mhi \leq 10^5 \, \Msol$ (i.e. below the lowest reported \hi mass for an observed dwarf galaxy). Our results thus predict the existence of isolated faint dwarfs that are temporarily hidden in \hinospace, as feedback-driven variability drives them below current observational capabilities of radio observatories. A clear observational signature to distinguish such temporarily hidden galaxies from long-lived quiescent objects is that they should exhibit young stars and evidence of recent star formation, as no galaxies in our star-forming sample stay without \hi for longer than 400 Myr. 

Quantifying how time variability will affect the overall population of \hinospace-bearing dwarf galaxies remains challenging at this stage, as our small sample size does not allow us to average over dependences on the specific star formation histories of our dwarfs. None the less, this strongly motivates dedicated modelling of such feedback-driven duty cycle, to account for potential detectability biases when interpreting constraints from the population of \hinospace-detected galaxies, but also as a promising discriminant between feedback models and their efficiency at driving outflows. Future studies with larger samples of low-mass dwarfs, and the ability to accurately track gas dynamics over short time-scales (e.g. \citealt{Genel2013, Cadiou2019}), will be key to achieving this aim.

\subsection{Spatial distribution of neutral gas} \label{sec:sec:histructure}

\begin{figure}
  \centering
    \includegraphics[width=\columnwidth]{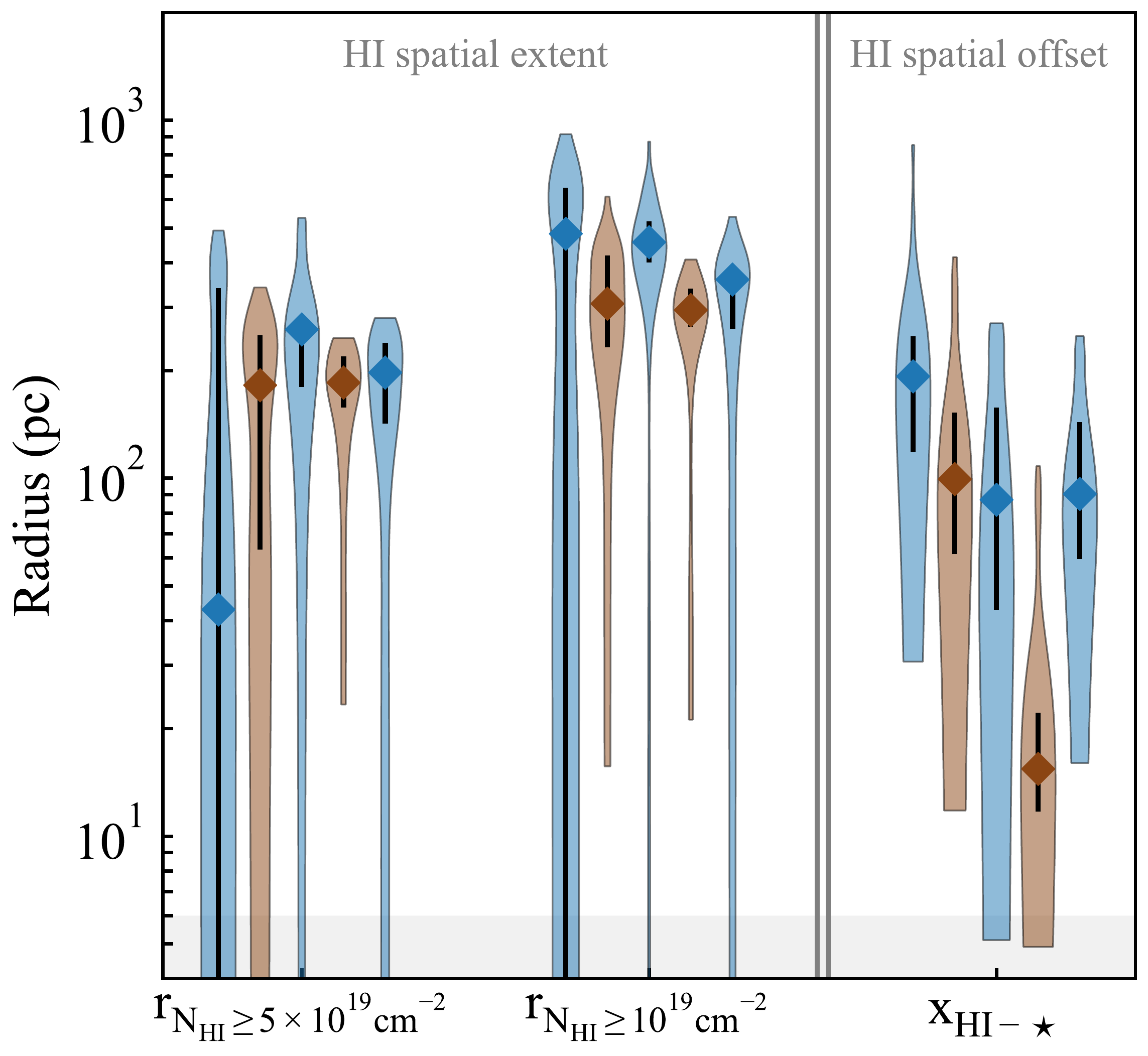}

    \caption{Time distribution of \hi structural parameters over the past four billion years in our simulated \hinospace-rich dwarfs (individual contours), showing the cylindrical radius enclosing depths of $\xScientific{5}{19}$ and $10^{19} \, \cmsquare$ (left and centre respectively) and the three-dimensional distance between \hi and stellar centres (right). Stellar feedback induces extended scatter in \hi spatial extents, with factor-of-a-few variations over time (IQR shown as black line). This variability is reduced with deeper \hi observations (black lines in left compared to centre) that better average spatial fluctuations, highlighting dependencies of the expected variability of observables on an experiment's depth. The same physical mechanism commonly generates offsets between the stellar and \hi components comparable to the galaxy's optical sizes ($\approx 200\, \pc$).
    }
    \label{fig:hisize}
\end{figure}

As illustrated in Figure~\ref{fig:himaps}, the spatial distribution of \hi undergoes large variations over time. We now turn to quantifying variability in the structural properties of \hi reservoirs. For each \hinospace-bearing galaxy, we compute their radial column density profile along a random line of sight\footnote{We will present kinematic analysis of the suite in a follow-up paper but find no preferential orientation or ordered rotation for our star-forming dwarfs, consistent with the observed population (\citealt{McQuinn2021}).} using 150 bins equally spaced in log space and determine the radius at which the profile drops below depths of $\xScientific{5}{19}$ and $10^{19} \, \cmsquare$ over the last four billion years. We show the distribution of values over time in Figure~\ref{fig:hisize} (left and centre). When the entire \hi distribution is fainter than a chosen depth cut, we associate a vanishing \hi size to this galaxy, creating the vertical tails in Figure~\ref{fig:hisize} extending below twice the maximum spatial resolution of the simulation (6 pc; shading).

There is qualitative agreement between the medians of our simulated galaxies and observed \hi sizes in galaxies of comparable stellar masses such as Leo T with $\rhiaboveXX{3}{19} \approx 400 \, \pc$ (\citealt{Adams2018}) and Leo P with $\rhiaboveXX{5}{19} \approx 300 \, \pc$ (\citealt{Giovanelli2013}). However, our results highlight that, similar to the variability in total \hi mass, such sizes are significantly time-varying in all galaxies, with 50 per cent confidence intervals extending from 0.14 to 0.60 dex around the medians at sensitivities $\Nhi \geq \xScientific{5}{19} \, \cmsquare$. We verified that time variations in spatial extents are positively correlated with those in neutral gas masses (i.e. larger \hi masses correlate with larger \hi extents; see also e.g. \citealt{Wang2016}), although with an extended scatter and thus driving surface-brightness fluctuations in 21 cm.

Ongoing and future radio surveys will likely reach depths closer to $\Nhi = \xScientific{1}{19} \, \cmsquare$ (e.g. \citealt{Maddox2021}) when probing a dwarf's \hi distribution. This systematically increases the median \hi extent of each galaxy as expected (middle columns), but also reduces variability with systematically smaller confidence intervals. (Conversely, restricting the extent to $\Nhi = \xScientific{1}{20} \, \cmsquare$ produces the opposite trend, with smaller medians and larger confidence intervals.) Shallower depths focus on a galaxy's centre, where most stirring due to feedback from old and young stars occurs. Deeper \hi observations reaching outwards spatially average perturbations in the galaxy's centre, leading to more stable sizes whose variability is then driven by larger, rarer events. 

Our results demonstrate that the level of variability in \hi surface brightnesses thus depends on depth, which has complex consequences for the detectability of a given dwarf. Modelling it requires accounting for the interplay between variability, an object's distance and a given survey design of sensitivity and angular resolution (see Appendix~\ref{app:resolution} for a visual illustration). Carefully quantifying how time variability affects completeness limits in a population of Local Volume, \hinospace-rich dwarfs is thus a necessary undertaking for future studies aiming to constrain cosmological galaxy formation through neutral gas observations and inform survey strategy through detection rates.

As well as modifying the size of the \hi distribution, stellar feedback also induces asymmetric morphologies and spatial offsets between the neutral gas and stellar components (Figure~\ref{fig:himaps}, see also e.g. \citealt{Read2016DwarfRCs, El-Badry2018}). We quantify this latter aspect in Figure~\ref{fig:hisize} (right-hand columns), reporting the median and IQR for the three-dimensional distance between the stellar and \hi centres over the last four billion years. All galaxies exhibit spatial offsets between their stars and neutral gas that are regularly comparable to their optical sizes ($\rhalflight \approx 200 \, \pc$, see Table~\ref{table:runs}). Feedback-driven outflows thus provide a natural explanation to overlapping, but offset, distributions of stellar and neutral gas material at a given time (as reported by e.g. \citealt{Janesh2019}). Our results further provide an alternative interpretation to environmental processing in explaining misaligned distributions and asymmetric morphologies within galaxies closer to us, where it remains unclear whether they are interacting with our Milky Way (e.g. Leo T; \citealt{Adams2018}).

\section{Discussion and conclusions} \label{sec:conclusion}

We have shown how future surveys should uncover diversity in the neutral gas observables of faint, low-mass dwarf galaxies ($\Mstar \lessapprox 10^7\, \Msol$). We assembled a suite of 10 simulated dwarf galaxies evolved to $z = 0$ using high-resolution zoomed cosmological simulations from the EDGE project (\citealt{Rey2019UFDScatter, Rey2020, Agertz2020EDGE, Orkney2021}), bracketing the transition from quiescent to gas-rich to star-forming low-mass dwarfs (\citealt{Rey2020}). We demonstrated how the interplay between dynamical mass assemblies and feedback introduces extended scatter in the relationship between \hi and stellar observables at the faint end (illustrated in Figure~\ref{fig:himaps}).
\begin{itemize}
  \item We re-affirm that cosmic reionization and photoionization feedback from the UVB plays a key role in truncating the faint end of the \hi mass function (e.g. \citealt{Efstathiou1992, Benitez-Llambay2017, Tollerud2018}), but show that this cut-off is not localized in galaxy stellar mass. Rather, the diversity of possible assemblies at early and late times leads to a bimodality at the faint end of the $\Mstar-\Mhi$ relation with \hinospace-bearing and \hinospace-deficient isolated dwarf galaxies cohabitating over an extended range of stellar masses (Figure~\ref{fig:himass}). 
  \item Stellar feedback within such shallow potential wells ($\Mvir \lessapprox \xMsol{3}{9}$) drives significant time variability in \hi observables, generating order-of-magnitude scatter around the median $\Mhi$ at fixed $\Mstar$ over the last four billion years (Figure~\ref{fig:himass}). Furthermore, star-forming dwarfs undergo regular, short-lived episodes without detectable neutral gas reservoirs following intense outflows. This leads to a feedback-driven duty cycle in low-mass dwarfs' \hi contents affecting their detectability by radio observatories. 
  \item Feedback also generates time-varying sizes and asymmetric morphologies in \hi spatial distributions (Figure~\ref{fig:hisize}). This implies regular surface-brightness fluctuations over time and offsets between \hi and stellar components, providing a natural explanation for disturbed and misaligned \hi morphologies in low-mass dwarfs. Such disturbed distributions make mass modelling -- either through rotation curve analysis or assuming steady-state hydrostatic equilibrium -- challenging in these galaxies, which will be the focus of a forthcoming study.
\end{itemize}

The physical mechanisms uncovered in this work have key consequences for future studies aiming to constrain reionization, the history of the UVB across cosmic time, and cosmological dwarf galaxy formation through the population of \hinospace-bearing systems (e.g. \citealt{Benitez-Llambay2017, Tollerud2018, Benitez-Llambay2020}). In particular, capturing the existence of potential starless \hi clouds and our exposed bimodality at the faint end of the $\MstarMhi$ relation requires modelling at once the growth of stellar mass in dark matter haloes pre- and post-reionization, while accounting for the interplay between the UVB and their dynamical mass assemblies setting their gas reservoirs at late times. Performing such an exercise on a large sample of dark matter haloes is thus highly desirable. In future work (Kim et al., in preparation), we plan to use a semi-analytic model of dwarf galaxy formation to predict the width of the bimodality at the faint end of the $\MstarMhi$.

In turn, the characterization of the relationship between stellar and \hi masses across a population of dwarfs could provide constraints on cosmic reionization complementary to more direct measures (e.g. \citealt{Tollerud2018}). Our results strongly motivate future studies aiming to perform this inference, but robust constraints are challenging at this stage. In addition to the interaction between mass growth and feedback exhibited by our study, the timing of reionization at high redshift for each given dwarf depends on their local environment (e.g. \citealt{Katz2020, Ocvirk2020}). This will introduce a modulation for their pre-reionization stellar masses independent of later evolution, whose impact on the $z=0$ $\MstarMhi$ relation will have to be quantified. Furthermore, uncertainties in the amount of heating at intermediate redshifts (due to helium reionization, $z\approx3$; e.g. \citealt{UptonSanderbeck2016}) and later times ($z\leq1$; e.g. \citealt{Gaikwad2017, Khaire2019Measurements}) affect the absolute halo mass capable of hosting an \hi dwarf at a given time. Such aspects will need to be explored by future studies aiming to robustly constrain reionization through the observed population of \hinospace-bearing dwarf galaxies.

Furthermore, modelling the scatter in the $\MstarMhi$ relation will require quantifying the expected variability in \hi of low-mass dwarfs due to feedback exhibited in this work. This will be a key undertaking for future studies due to its implications for the detectability of low-mass dwarfs at a given time. A duty cycle of the overall \hi content, combined with our exposed surface-brightness fluctuations in 21 cm emission, will be necessary to account for when interpreting completeness estimates of observed populations of \hinospace-bearing dwarfs and predicting detection rates to inform survey strategy for future experiments (e.g. the Square Kilometer Array). 

A key aspect to achieve this aim will be to understand the relationship between variability and the expected burstiness of star formation in such shallow potential wells. Despite the resolution of our simulations and our accurate modelling of supernova explosions, uncertainties remain in their coupling with the dwarf's surrounding ISM and subsequent efficiency in driving outflows (\citealt{Smith2019, Agertz2020EDGE, Smith2020PhotoRT}). In particular, the addition of feedback channels such as photoionization feedback can lead to more gentle, less explosive regulation of star formation (e.g. \citealt{Agertz2020EDGE, Smith2020PhotoRT}). Re-simulating all dwarfs accounting for radiative effects is beyond the scope of this work and will be tackled in future work (Taylor et al., in preparation), but we provide an illustration of the sensitivity of \hi variability to feedback models. We repeat our analysis on a re-simulation of our prototypical star-forming, gas-rich dwarf (Halo 600) artificially reducing the strength of supernova feedback (`Weak feedback' model in \citealt{Agertz2020EDGE}, first presented in \citealt{Rey2020}). This results in an unphysically high galaxy stellar mass (from $\Mstar= \xMsol{3.5}{5}$ to $\xMsol{1.5}{7}$ at $z=0$) at this halo mass (\citealt{Read2017}), but demonstrates that less efficient outflows suppress variability in \hi mass (IQR going from 1.83 to 0.13 dex around the median). With this model, the galaxy also never undergoes `unobservable' episodes with $\Mhi < \xMsol{1}{5}$. Pinpointing these uncertainties, from additional feedback channels (e.g. \citealt{Agertz2020EDGE, Smith2020PhotoRT}) or physical alterations to the feedback budget (e.g. \citealt{Prgomet2021}), will thus be critical.

Finally, this demonstrated sensitivity to feedback models also motivates future studies aiming to constrain the \hi duty cycle from an observed population of dwarfs. Constraining such duty cycle and variability from data bears strong similarities to the modelling of variable active galactic nuclei populations, that provides constraints on their luminosity over time and the available feedback budget to couple to their host galaxies (e.g. \citealt{Wyithe2002, Conroy2013, Delvecchio2020}). Such statistical measurement would be an ideal complement to resolved, multi-wavelength studies of faint dwarfs (\citealt{McQuinn2019Outflows, Zheng2020}) to inform on the prevalence and mass-loading factors of outflows in the smallest star-forming galaxies. 

\section*{Acknowledgements}
We thank the referee for a constructive review that improved the quality of the manuscript. MR thanks Eric Andersson, Anastasia Ponomareva, Lorenzo Posti, and Florent Renaud for insightful discussions during the construction of this work. MR would like to further thank Claire H\'ebert, Antoine Petit and Hans for their hospitality and company during a trip in which most of this work was conceived. MR is supported by the Beecroft Fellowship funded by Adrian Beecroft. MR and OA acknowledge support from the Knut and Alice Wallenberg Foundation, the Swedish Research Council (grants 2014-5791 and 2019-04659) and the Royal Physiographic Society in Lund. AP is supported by the Royal Society. MO acknowledges the UKRI Science and Technology Facilities Council (STFC) for support (grant ST/R505134/1). This project has received funding from the European Union’s Horizon 2020 research and innovation programme under grant agreement No. 818085 GMGalaxies. This work was performed using the DiRAC Data Intensive service at Leicester, operated by the University of Leicester IT Services, which forms part of the STFC DiRAC HPC Facility (www.dirac.ac.uk). The equipment was funded by BEIS capital funding via STFC capital grants ST/K000373/1 and ST/R002363/1 and STFC DiRAC Operations grant ST/R001014/1. DiRAC is part of the National e-Infrastructure. We acknowledge the use of the UCL Grace High Performance Computing Facility, the Surrey Eureka supercomputer facility, and associated support services. This work was partially supported by the UCL Cosmoparticle Initiative.

\section*{Author contributions}
The main roles of the authors were, using the CRediT (Contribution Roles Taxonomy) system\footnote{\url{https://authorservices.wiley.com/author-resources/Journal-Authors/open-access/credit.html}}:

MR: Conceptualisation ; Data curation; Formal analysis; Investigation; Writing – original draft. AP: Conceptualisation; Funding Acquisition; Methodology; Writing – review and editing. OA: Funding Acquisition; Methodology; Software; Writing – review and editing. MO: Data Curation; Writing – review and editing. JR: Conceptualisation; Funding Acquisition; Project Administration; Resources. Writing – review and editing. AS: Writing – review and editing. SK: Conceptualisation. PD: Writing – review and editing.

\section*{Data Availability}
The data underlying this paper will be shared on reasonable request to the corresponding author.

%%%%%%%%%%%%%%%%%%%%%%%%%%%%%%%%%%%%%%%%%%%%%%%%%%



\bibliographystyle{mnras}
\bibliography{HI_Bilbio} 

%%%%%%%%%%%%%%%%%%%%%%%%%%%%%%%%%%%%%%%%%%%%%%%%%%

%%%%%%%%%%%%%%%%% APPENDICES %%%%%%%%%%%%%%%%%%%%%

\appendix
\section{The coupling between time variability and spatial resolution} \label{app:resolution}

\begin{figure*}
  \centering
    \includegraphics[width=\textwidth]{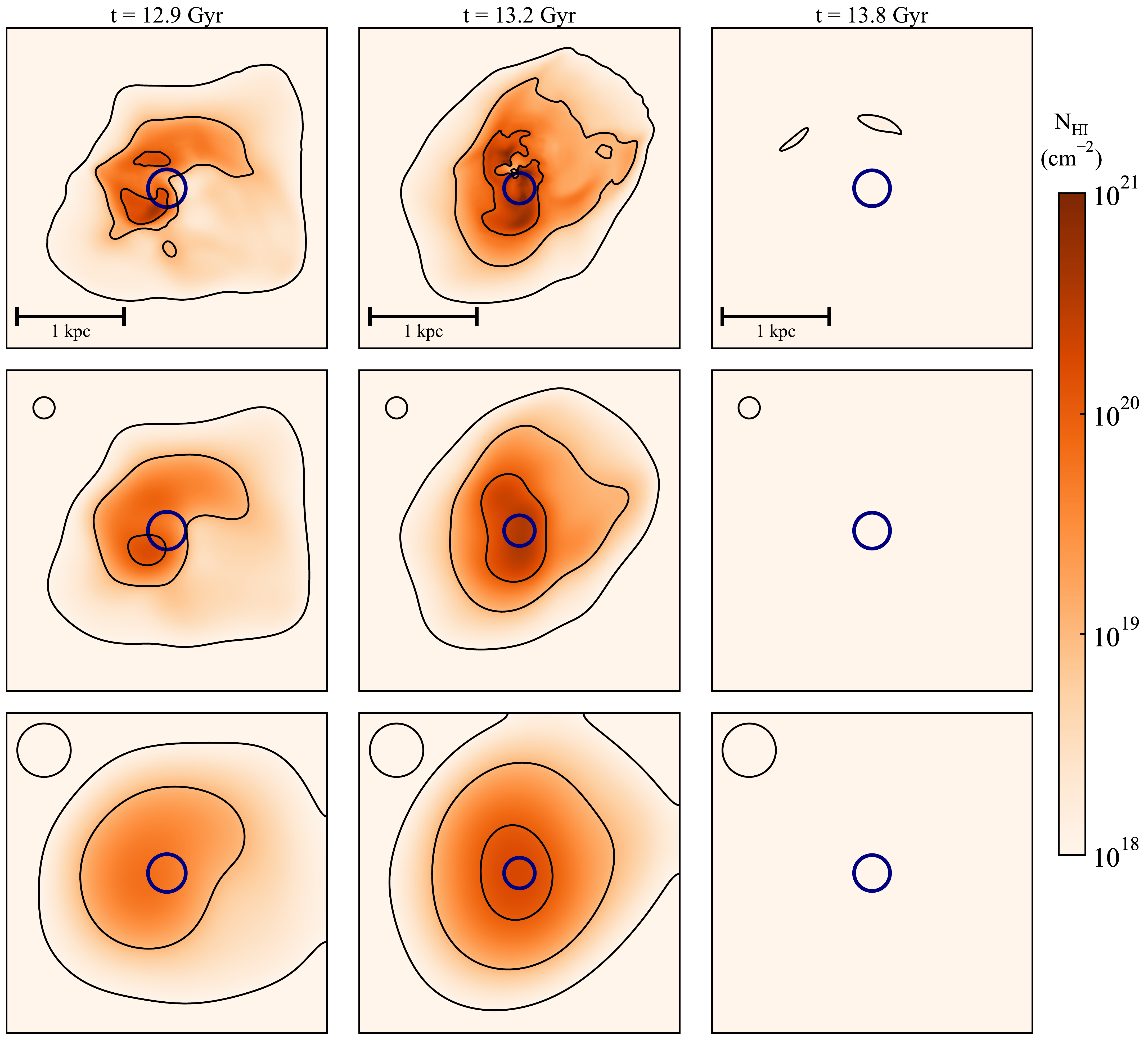}

    \caption{Illustrating how different spatial resolutions combine with time variability to affect a dwarf's detectability. We show the same \hi column density maps from our example `star-forming, gas-rich' dwarf in Figure~\ref{fig:himaps} (top row) convolved by a Gaussian beam of 200 and 500 pc (middle and bottom rows). As spatial resolution is degraded, \hi emission becomes more diffuse and symmetric, eventually fading out the dwarf and its features (individual columns). This effect couples with feedback-driven surface-brightness fluctuations (Section~\ref{sec:results}, individual lines), which can make a dwarf challenging to observe because its \hi reservoir has been temporarily removed (e.g. bottom right) or because \hi emission is too diffuse at the available spatial resolution (bottom left).}
    \label{fig:himapsmoothed}
\end{figure*}

In Figure~\ref{fig:himaps}, we show \hi column densities maps derived on a scale close to our simulations' resolution (6 pc) to showcase the full extent of the simulated data. However, the achieved spatial resolution by a given instrument depends on its angular resolution and the size of the observed object on the sky. A population of faint dwarfs spawning a distribution of distances will thus be observed at varying spatial resolutions, while feedback-driven variability leads to surface-brightness fluctuations in \hi emission (Section~\ref{sec:sec:histructure}). In this appendix, we provide a visual illustration of how these two effects combine, highlighting the importance of modelling both when predicting the observability of a faint \hi dwarf.

Figure~\ref{fig:himapsmoothed} shows column density maps from our `star-forming, gas-rich' dwarf at the same time snapshots as in Figure~\ref{fig:himaps} (top row reproduced between figures). We then convolve these images with a Gaussian beam with 200 and 500 pc diameter (middle and bottom row, respectively), and show contours of constant $10^{18}$, $10^{19}$, and $10^{20} \cmsquare$ \hi column densities in black. As expected, \hi emission becomes more diffuse and symmetric with decreasing spatial resolution (top to bottom in each individual column).

Comparing between columns in Figure~\ref{fig:himapsmoothed} illustrates the coupling between time variability and spatial resolution in affecting a dwarf's detectability. A sizeable \hi reservoir at high spatial resolution (top left) might remain detectable at medium resolution (centre left), showcasing a contour with $\NhiaboveXX{1}{20}$ offset but overlapping with $\rhalflight$ (blue circle), but would be challenging to detect at low angular resolution (column densities always below $\NhibelowXX{5}{19}$; bottom left). By contrast, the same object 300 Myr later (middle column) would likely be detectable at all spatial resolutions, and in turn undetectable at all resolutions a further 600 Myr later (right-hand column). The duty cycle of `observable' episodes in \hi for a given dwarf thus depends on the available spatial resolution. Accounting for these effects will be key to interpret completeness in the census of low-mass dwarfs from \hi surveys, and to inform observational strategies in future surveys.

%%%%%%%%%%%%%%%%%%%%%%%%%%%%%%%%%%%%%%%%%%%%%%%%%

% Don't change these lines
\bsp	% typesetting comment
\label{lastpage}
\end{document}